# The principle of microreversibility and the fluctuation relations for quantum systems driven out of equilibrium

Hiroshi Matsuoka

*Department of Physics, Illinois State University, Normal, Illinois, 61790-4560, USA*

Phone: (309) 438-3236

FAX: 309-438-5413

hmb@phy.ilstu.edu

## ABSTRACT

For classical systems driven out of equilibrium, Crooks derived a relation (the Crooks-Jarzynski relation), whose special cases include a relation (the Crooks relation) equivalent to the Kawasaki non-linear response relation. We derive a quantum extension of the Crooks-Jarzynski relation without explicitly using the principle of microreversibility. Its special cases lead to the Jarzynski equality and the standard linear response theory with a Green-Kubo formula with a canonical correlation function. We also derive a quantum extension of the Crooks relation using the principle of microreversibility. Its special cases lead to the Jarzynski equality, the Crooks transient fluctuation theorem, and the fluctuation theorem for current or shear stress, which leads to a Green-Kubo formula with a symmetrized correlation function. For each quantum Crooks relation, there exists a corresponding quantum Crooks-Jarzynski relation. Using either relation, we can derive the Jarzynski equality, the fluctuation theorems mentioned above, and the standard linear response theory.

## 1. INTRODUCTION AND A SUMMARY OF OUR RESULTS

Since 1993, a number of exact relations including the fluctuation theorems [1 – 12] and the Jarzynski equality [13] have been derived originally for classical systems driven out of equilibrium and later for quantum systems [14, 15]. In this article, we will call these exact relations the fluctuation relations.

For classical systems, Crooks [16] derived a relation (*i.e.*, Eq. (15) in [16]), which we will call the Crooks relation in this article, from what is equivalent to the detailed fluctuation theorem [9] and showed that its special cases include the Crooks transient fluctuation theorem [6], the Jarzynski equality, and a relation (*i.e.*, Eq. (21)[1] in [16]) equivalent to the Kawasaki non-linear response relation [17 – 21]. In this article, we will call this relation equivalent to the Kawasaki non-linear response relation the Crooks-Jarzynski relation.

For a classical system driven by an external field, Hayashi and Sasa [21] used the Kawasaki non-linear response relation[2] to recover the standard linear response theory [22] including the Green-Kubo relation for the conductivity of the system. For a classical system driven by a chemical potential difference between two particle reservoirs attached to the system, they also derived a relation equivalent to the Crooks relation and recovered the Green-Kubo relation for the conductivity of the system obtained earlier by the fluctuation theorem for current [4, 7, 8].

Using the Crooks relation for classical systems driven out of equilibrium, we can therefore derive the Crooks transient fluctuation theorem, the Jarzynski equality, the standard linear response theory, and the Green-Kubo relation obtained by the fluctuation theorem for current.

---

[1] In [16], Crooks acknowledged that this relation was also due to Jarzynski.
[2] Hayashi and Sasa used the relation that Crooks in [16] called the Kawasaki non-linear response relation although they reserved the term "the Kawasaki non-linear response relation" for the form of the relation closer to its original form [17 – 20].

For an isolated quantum system subject to a time-dependent external field, using the principle of microreversibility [14, 15], Bochkov and Kuzovlev [23] derived a general relation whose special cases include a relation[3] that Andrew and Gaspard [24] derived from their universal quantum work relation based on the principle of microreversibility. Using this relation, Andrew and Gaspard also recovered the standard linear response theory [22] with the Green-Kubo formula for a linear response coefficient in terms of a canonical correlation function of an operator corresponding to a quantity induced by the external field.

The Jarzynski equality for a quantum system evolving in time according to a time-dependent Hamiltonian was also derived from a special case of this relation [25, 26].

In this article, we will show, in Sec. 3.3, that this relation is a special case of a quantum extension of the Crooks-Jarzynski relation, which relates the forward statistical average of a product of operators including an arbitrary operator $\hat{A}$ and the canonical ensemble density matrices, $\hat{\rho}_{\text{in}}$ and $\hat{\rho}_{\text{fin}}$, for the initial and the final eigenstates of a system to the time-reversed backward statistical average of an operator $^{\Theta}\hat{A}$ defined by $^{\Theta}\hat{A} \equiv \hat{\Theta}\hat{A}\hat{\Theta}^{\dagger}$, where $\hat{\Theta}$ is the time reversal operator.

In Sec. 3.1, we will derive this quantum Crooks-Jarzynski relation without explicitly using the principle of microreversibility, which is consistent with the fact that the principle of microreversibility was not used explicitly in the original derivations of the standard linear response theory [22] and the Jarzynski equality [25, 26].

The quantum Crooks-Jarzynski relation derived in Sec. 3.1 holds only for an initial density matrix that is *positive-definite*, which is the case for the canonical ensemble density matrix defined in Sec. 2.2 and assumed in Sec. 3.1. In Sec. 3.4, we will then show that for a general initial density matrix that is *positive-*

---

[3] (3.22) in Sec. 3.3.

*semidefinite* but not necessarily *positive-definite*, such as a microcanonical ensemble density matrix, we can also derive the quantum Crooks-Jarzynski relation for a class of operators that satisfy a certain condition.

For a quantum system in a steady state with a constant current of heat or particles driven by a temperature or chemical potential difference between two reservoirs attached to the system, both the Crooks transient fluctuation theorem and the fluctuation theorem for the current were derived from the principle of microreversibility [27 – 29] and the fluctuation theorem for the current has been shown to lead to the Green-Kubo formula for a conductivity in terms of a symmetrized correlation function of the current density operator.

The fluctuation theorem for the heat current was also shown [30] to follow from a quantum extension of the Crooks relation, which relates the forward process average of a product of quantities including the canonical ensemble statistical distributions, $\rho_{in}$ and $\rho_{fin}$, for the initial and the final eigenstates of the system and a complex-valued quantity $C(i,f)$, which may depend on the initial and the final eigenstates of a system.

For a quantum system in a steady state with a constant rate of work done on the system such as a fluid in a steady shear flow, the fluctuation theorem for the quantity induced in the system, such as the shear stress on the fluid, was derived [30] from a special case[4] of the quantum Crooks relation and was shown to lead to the Green-Kubo formula for a linear response coefficient such as the shear viscosity of the fluid in terms of a symmetrized correlation function of an operator corresponding to the induced quantity.

In Sec. 4.1, using a direct consequence[5] of the principle of microreversibility, we will derive a general form of the quantum Crooks relation, which relates the forward process average of a product of quantities including statistical distributions, $\rho_{\text{in}}$ and $\rho_{\text{fin}}$, for the initial and the final eigenstates of the system and a complex-valued quantity $C(i,f)$, which may depend on the initial and the final eigenstates of a system and vanishes for an initial eigenstate for which $\rho_{\text{in}}$ vanishes, to the time-reversed backward process average of the same quantity $C(i,f)$.

---

[4] (4.25) in Sec. 4.4.
[5] (2.13) in Sec. 2.4.

In Sec. 4.3, we will also derive the Crooks transient fluctuation theorem from the quantum Crooks relation. As we will mention at the end of Sec. 3.2, the Jarzynski equality is also a special case of the quantum Crooks relation.

For quantum systems driven out of equilibrium, it therefore appears that we can derive the standard linear response theory and the Jarzynski equality directly from the quantum Crooks-Jarzynski relation without explicitly using the principle of microreversibility while we need to use the quantum Crooks relation based on the principle of microreversibility to derive the Crooks transient fluctuation theorem and the fluctuation theorem for the current or shear stress.

A natural question is then how exactly the quantum Crooks-Jarzynski relation and the quantum Crooks relation are related with each other. In Sec. 3.4, we will show that for each quantity $C(i,f)$ that will be later shown to satisfy a quantum Crooks relation in Sec. 4.1, we can define a corresponding operator $\hat{A}_C$ that satisfies a quantum Crooks-Jarzynski relation. In Sec. 4.2, we will then show that the quantum Crooks relation for $C(i,f)$ and quantum Crooks-Jarzynski relation for $\hat{A}_C$ are in fact equivalent to each other so that using this corresponding quantum Crooks-Jarzynski relation for $\hat{A}_C$, we can derive any result that follows from the quantum Crooks relation for $C(i,f)$ and vice versa. More specifically, using the quantum Crooks-Jarzynski relation, we can then derive the Crooks transient fluctuation theorem and the fluctuation theorem for the current or shear stress.

In Sec. 4.2, we will also note that a specific case of quantum Crooks relation corresponds to the special case of quantum Crooks-Jarzynski relation that leads to the standard linear response theory so that using this special case of quantum Crooks relation, we can also derive the standard linear response theory.

Therefore, using either the quantum Crooks-Jarzynski relation or the quantum Crooks relation, we can derive the fluctuation relations mentioned above and the standard linear response theory.

Both the general quantum Crooks relation and its corresponding quantum Crooks-Jarzynski relation are quite general mathematical identities and neither the statistical distributions $\rho_{in}$ and $\rho_{fin}$ involved in the general quantum Crooks relation nor the density matrices $\hat{\rho}_{in}$ and $\hat{\rho}_{fin}$ involved in the corresponding quantum Crooks-Jarzynski relation need to represent equilibrium ensembles. This suggests that we may be able to apply these relations to quantum systems evolving from a non-equilibrium state to another.

## 2. SYSTEM AND ITS FORWARD AND TIME-REVERSED BACKWARD PROCESSES

### 2.1. System Hamiltonian and the time evolution operator for a forward process

For simplicity, we will consider an isolated quantum system that evolves, during a time interval $[0,\tau]$, according to a time-dependent Hamiltonian $\hat{H}(t)$. It is straightforward to extend our discussion in this article to systems attached to heat or particle reservoirs as we can regard a composite system consisting of such a system and reservoirs as an isolated system evolving in time according to a total

Hamiltonian that includes an interaction term between the system and the reservoirs.

During a forward process over the time interval $[0,\tau]$, the state $|\Psi(t)\rangle$ of the system evolves according to the Schrödinger equation with $\hat{H}(t)$ so that its final state is related to its initial state by

$$|\Psi(\tau)\rangle = \hat{U}(0,\tau)|\Psi(0)\rangle, \qquad (2.1)$$

where $\hat{U}(0,\tau)$ is the time evolution operator for the forward process during $[0,\tau]$. In the rest of this article, $\hat{U}(0,\tau)$ will be simply denoted by $\hat{U}$.

### 2.2. Initial energy eigenstates of the forward process

Just before the initial time $t=0$, through a measurement of the energy in the system, we find the system to be in an eigenstate $|i\rangle$ of its initial Hamiltonian $\hat{H}(0)$ with energy eigenvalue $E(i)$:

$$\hat{H}(0)|i\rangle = E(i)|i\rangle. \qquad (2.2)$$

Before $t=0$, we assume that the system is in its equilibrium state at an inverse temperature $\beta$ so that the initial eigenstate $|i\rangle$ is selected by the following canonical ensemble distribution:

$$\rho_{in}(i) \equiv \frac{1}{Z_{in}(\beta)} \exp[-\beta E(i)] = \exp(\beta F_{in})\exp[-\beta E(i)], \qquad (2.3)$$

where $Z_{in}$ defined by $Z_{in}(\beta) \equiv \sum_i \exp[-\beta E(i)]$ is the partition function for the system and is related to the Helmholtz free energy $F_{in}$ of the system in its initial equilibrium state by $Z_{in} = \exp(-\beta F_{in})$.

We then define the density matrix for the system corresponding to $\rho_{in}$ by

$$\hat{\rho}_{in} \equiv \frac{1}{Z_{in}(\beta)} \exp[-\beta \hat{H}(0)] = \exp(\beta F_{in}) \exp[-\beta \hat{H}(0)], \quad (2.4)$$

which satisfies

$$\hat{\rho}_{in} |i\rangle = \rho_{in}(i) |i\rangle. \quad (2.5)$$

### 2.3. Final energy eigenstates of the forward process

Just after the final time $t = \tau$, through a measurement of the energy in the system, we find the system to be in an eigenstate $|f\rangle$ of its final Hamiltonian $\hat{H}(\tau)$ with energy eigenvalue $E(f)$:

$$\hat{H}(\tau) |f\rangle = E(f) |f\rangle. \quad (2.6)$$

We assume that the set of all the initial eigenstates as well as the set of all the final eigenstates can serve as an orthonormal basis for the Hilbert space for the system so that we can use $\sum_i |i\rangle\langle i| = I$ and $\sum_f |f\rangle\langle f| = I$.

### 2.4. Time-reversed backward process and the principle of microreversibility

In this section, to make this article to be self-contained, we will derive the principle of microreversibility (2.12) and its direct consequence (2.13). The readers who are familiar with these equations may wish to skip this subsection.

As the time reversal operator $\hat{\Theta}$ satisfies $i\hat{\Theta} = -\hat{\Theta}i$ and $\hat{\Theta}\hat{\Theta}^{\dagger} = \hat{\Theta}^{\dagger}\hat{\Theta} = I$, the time-reversed state defined by

$$|\Psi_r(t)\rangle \equiv \hat{\Theta}|\Psi(\tau - t)\rangle \qquad (2.7)$$

evolves according to the following Schrödinger equation:

$$\begin{aligned}
i\hbar\frac{\partial}{\partial t}|\Psi_r(t)\rangle &= i\hbar\frac{\partial}{\partial t}\left(\hat{\Theta}|\Psi(\tau-t)\rangle\right) = \hat{\Theta}\left\{i\hbar\frac{\partial}{\partial(\tau-t)}|\Psi(\tau-t)\rangle\right\} \\
&= \hat{\Theta}\hat{H}(\tau-t)|\Psi(\tau-t)\rangle = \hat{\Theta}\hat{H}(\tau-t)\hat{\Theta}^{\dagger}\left(\hat{\Theta}|\Psi(\tau-t)\rangle\right) \\
&= {}^{\Theta}\hat{H}(\tau-t)|\Psi_r(t)\rangle,
\end{aligned}$$

$$(2.8)$$

where the time-reversed Hamiltonian ${}^{\Theta}\hat{H}(\tau-t)$ is defined by

$$^{\Theta}\hat{H}(\tau-t) \equiv \hat{\Theta}\hat{H}(\tau-t)\hat{\Theta}^{\dagger}. \qquad (2.9)$$

The final state of the time-reversed backward process is related to its initial state by

$$|\Psi_r(\tau)\rangle = {}^{\Theta}\hat{U}(0,\tau)|\Psi_r(0)\rangle, \qquad (2.10)$$

where ${}^{\Theta}\hat{U}(0,\tau)$ is the time evolution operator for the backward process, which is controlled by ${}^{\Theta}\hat{H}(\tau-t)$. In the rest of this article, ${}^{\Theta}\hat{U}(0,\tau)$ will be simply denoted by ${}^{\Theta}\hat{U}$.

For any $|\Psi(0)\rangle$, we then find

$$\hat{\Theta}|\Psi(0)\rangle = |\Psi_r(\tau)\rangle = {}^\Theta\hat{U}|\Psi_r(0)\rangle = {}^\Theta\hat{U}\hat{\Theta}|\Psi(\tau)\rangle = {}^\Theta\hat{U}\hat{\Theta}\hat{U}|\Psi(0)\rangle \qquad (2.11)$$

so that

$$^\Theta\hat{U} = \hat{\Theta}\hat{U}^\dagger\hat{\Theta}^\dagger, \qquad (2.12)$$

which is called the principle of microreversibility. The principle of microreversibility therefore relates the time evolution operator $^\Theta\hat{U}$ for the backward process to the time evolution operator $\hat{U}$ for the forward process and it is a general property of the time evolution operator $^\Theta\hat{U}$ for any quantum system.

Using (2.12), we can also show that the transition probability for the forward process from an initial eigenstate $|i\rangle$ to a final eigenstate $|f\rangle$ is equal to the transition probability for the backward time-reversed process from $|^\Theta f\rangle \equiv \hat{\Theta}|f\rangle$ to $|^\Theta i\rangle \equiv \hat{\Theta}|i\rangle$:

$$\left|\langle f|\hat{U}|i\rangle\right|^2 = \left|\langle ^\Theta i|^\Theta\hat{U}|^\Theta f\rangle\right|^2, \qquad (2.13)$$

which follows from

$$\langle ^\Theta i|^\Theta\hat{U}|^\Theta f\rangle = \left(\hat{\Theta}|i\rangle,\ \hat{\Theta}\hat{U}^\dagger|f\rangle\right) = \left(\hat{U}^\dagger|f\rangle,\ |i\rangle\right) = \langle f|\hat{U}|i\rangle, \qquad (2.14)$$

where $\hat{\Theta}$ is anti-unitary so that for any pair of states, $|\alpha\rangle$ and $|\alpha'\rangle$, $\hat{\Theta}$ satisfies

$$\left(\hat{\Theta}|\alpha'\rangle,\ \hat{\Theta}|\alpha\rangle\right) = \left(|\alpha\rangle,\ |\alpha'\rangle\right), \qquad (2.15)$$

where $\left(|\alpha\rangle,\ |\alpha'\rangle\right)$ is the inner product between $|\alpha\rangle$ and $|\alpha'\rangle$.

(2.13) is also a general property of the time evolution operator $^{\Theta}\hat{U}$ for any quantum system and we will use it in Sec. 4.1, where we will derive the quantum Crooks relation.

## 2.5. A useful property of the time reversal operator

The following property of $\hat{\Theta}$ will be also useful in Sec. 4. If $|n\rangle$ is an eigenstate of an observable $\hat{O}$ with a real eigenvalue $a(n)$ so that $\hat{O}|n\rangle = a(n)|n\rangle$, then $|^{\Theta}n\rangle \equiv \hat{\Theta}|n\rangle$ is an eigenstate of $^{\Theta}\hat{O} \equiv \hat{\Theta}\hat{O}\hat{\Theta}^{\dagger}$ with the eigenvalue $a(n)$:

$$^{\Theta}\hat{O}|^{\Theta}n\rangle = \hat{\Theta}\hat{O}\hat{\Theta}^{\dagger}\hat{\Theta}|n\rangle = \hat{\Theta}\hat{O}|n\rangle = a(n)\hat{\Theta}|n\rangle = a(n)|^{\Theta}n\rangle. \qquad (2.16)$$

## 3. THE QUANTUM CROOKS-JARZYNSKI RELATION

### 3.1. Derivation of the quantum Crooks-Jarzynski relation

In this section, we will derive the quantum extension of the Crooks-Jarzynski relation, which relates the forward statistical average of a product of operators including an arbitrary operator $\hat{A}$ to the backward statistical average of an operator $^{\Theta}\hat{A}$ defined by

$$^{\Theta}\hat{A} \equiv \hat{\Theta}\hat{A}\hat{\Theta}^{\dagger}. \qquad (3.1)$$

The forward statistical average $\langle\hat{C}\rangle_{\mathrm{F}}$ of an operator $\hat{C}$ is defined by

$$\langle\hat{C}\rangle_{\mathrm{F}} \equiv \mathrm{Tr}[\hat{\rho}_{\mathrm{in}}\hat{C}] = \sum_{i}\langle i|\hat{\rho}_{\mathrm{in}}\hat{C}|i\rangle = \sum_{\substack{i \\ \rho_{\mathrm{in}}(i) \neq 0}} \rho_{\mathrm{in}}(i)\langle i|\hat{C}|i\rangle \qquad (3.2)$$

while the backward statistical average $\langle \hat{C} \rangle_R$ of an operator $\hat{C}$ is defined by

$$\langle \hat{C} \rangle_R \equiv \text{Tr}\left[ {}^\Theta \hat{\rho}_{\text{fin}} \hat{C} \right]. \qquad (3.3)$$

${}^\Theta \hat{\rho}_{\text{fin}}$ is defined by

$$ {}^\Theta \hat{\rho}_{\text{fin}} \equiv \hat{\Theta} \hat{\rho}_{\text{fin}} \hat{\Theta}^\dagger, \qquad (3.4)$$

where $\hat{\rho}_{\text{fin}}$ is a density matrix for the final eigenstates of the system and satisfies

$$\text{Tr}[\hat{\rho}_{\text{fin}}] = 1. \qquad (3.5)$$

The quantum Crooks-Jarzynski relation is then the following general identity for $\hat{A}$:

$$\left\langle \hat{A}_F(\tau) (\hat{U}^\dagger \hat{\rho}_{\text{fin}} \hat{U}) \frac{1}{\hat{\rho}_{\text{in}}} \right\rangle_F = \langle {}^\Theta \hat{A} \rangle_R, \qquad (3.6)$$

where $\hat{A}_F(\tau)$ is defined by

$$\hat{A}_F(\tau) \equiv \hat{U}^\dagger \hat{A} \hat{U}. \qquad (3.7)$$

We can show (3.6) as follows.

$$\left\langle \hat{A}_F(\tau) (\hat{U}^\dagger \hat{\rho}_{\text{fin}} \hat{U}) \frac{1}{\hat{\rho}_{\text{in}}} \right\rangle_F = \text{Tr}\left[ \hat{\rho}_{\text{in}} (\hat{U}^\dagger \hat{A} \hat{U})(\hat{U}^\dagger \hat{\rho}_{\text{fin}} \hat{U}) \frac{1}{\hat{\rho}_{\text{in}}} \right] = \text{Tr}[\hat{\rho}_{\text{fin}} \hat{A}]$$
$$= \text{Tr}[\hat{\Theta} \hat{\rho}_{\text{fin}} \hat{\Theta}^\dagger \hat{\Theta} \hat{A} \hat{\Theta}^\dagger] = \text{Tr}[{}^\Theta \hat{\rho}_{\text{fin}} {}^\Theta \hat{A}] = \langle {}^\Theta \hat{A} \rangle_R,$$

$$(3.8)$$

where we have used $\hat{U}\hat{U}^\dagger = I$, $\hat{\Theta}^\dagger\hat{\Theta} = I$, and the following property of the trace: $\text{Tr}[\hat{B}\hat{C}\hat{D}] = \text{Tr}[\hat{C}\hat{D}\hat{B}] = \text{Tr}[\hat{D}\hat{B}\hat{C}]$.

The operator $1/\hat{\rho}_{in}$ appearing in the right-hand side of the quantum Crooks-Jarzynski relation is well-defined because the initial density matrix $\hat{\rho}_{in}$ defined by (2.4) is a canonical ensemble density matrix, which is *positive-definite*. In Sec. 3.4, we will also show that for a general initial density matrix that is always *positive-semidefinite* but not necessarily *positive-definite*, such as a microcanonical ensemble density matrix, we can also derive the quantum Crooks-Jarzynski relation for a class of operators that satisfy a certain condition.

Note that we can derive the quantum Crooks-Jarzynski relation without explicitly using the principle of microreversibility, (2.12). As this derivation shows, the quantum Crooks-Jarzynski relation is simply based on $\hat{U}\hat{U}^\dagger = I$, $\hat{\Theta}^\dagger\hat{\Theta} = I$, and the property of the trace. It is rather remarkable that such a deceptively simple relation leads to highly nontrivial results such as the Jarzynski equality (see Sec. 3.2) and the standard linear response theory with the Green-Kubo formula (see Sec. 3.3).

Note also that the quantum Crooks-Jarzynski relation is a general mathematical identity that holds for any arbitrary operator $\hat{A}$ and arbitrary density matrices $\hat{\rho}_{in}$ and $\hat{\rho}_{fin}$ as long as $\hat{\rho}_{in}$ is positive-definite. Furthermore, the operator $\hat{A}$ does not need to be a Hermitian observable.

In addition, the density matrix $\hat{\rho}_{in}$ or $\hat{\rho}_{fin}$ does not need to represent an equilibrium ensemble for the initial or the final eigenstates of the system, which suggests that we may apply this quantum Crooks-Jarzynski relation to non-equilibrium ensembles that are represented by some density matrices as long as the initial density matrix is positive-definite.

For classical systems, Crooks [16] showed that the Crooks-Jarzynski relation is equivalent to the Kawasaki non-linear response relation. Within the framework of quantum dynamics, where the principle of microreversibility, ${}^{\Theta}\hat{U} = \hat{\Theta}\hat{U}^{\dagger}\hat{\Theta}^{\dagger}$, (2.12) always holds, the quantum Crooks-Jarzynski relation is also equivalent to the following quantum extension of the Kawasaki non-linear response relation:

$$\left\langle \hat{A}_F(\tau) \right\rangle_F = \left\langle {}^{\Theta}\hat{A}\left({}^{\Theta}\hat{U}^{\dagger\Theta}\hat{\rho}_{in}{}^{\Theta}\hat{U}\right)\frac{1}{{}^{\Theta}\hat{\rho}_{fin}} \right\rangle_R, \qquad (3.9)$$

as we can derive this relation using the quantum Crooks-Jarzynski relation (3.6) and the principle of microreversibility (2.12):

$$\begin{aligned}
\left\langle \hat{A}_F(\tau) \right\rangle_F &= \left\langle \left\{\hat{A}_F(\tau)\hat{\rho}_{in}\left(\hat{U}^{\dagger}\frac{1}{\hat{\rho}_{fin}}\hat{U}\right)\right\}\left(\hat{U}^{\dagger}\hat{\rho}_{fin}\hat{U}\right)\frac{1}{\hat{\rho}_{in}} \right\rangle_F \\
&= \left\langle \hat{U}^{\dagger}\left\{\hat{A}\hat{U}\hat{\rho}_{in}\left(\hat{U}^{\dagger}\frac{1}{\hat{\rho}_{fin}}\right)\right\}\hat{U}\left(\hat{U}^{\dagger}\hat{\rho}_{fin}\hat{U}\right)\frac{1}{\hat{\rho}_{in}} \right\rangle_F \\
&= \left\langle {}^{\Theta}\left\{\hat{A}\hat{U}\hat{\rho}_{in}\left(\hat{U}^{\dagger}\frac{1}{\hat{\rho}_{fin}}\right)\right\} \right\rangle_R = \left\langle \hat{\Theta}\left\{\hat{A}\hat{U}\hat{\rho}_{in}\left(\hat{U}^{\dagger}\frac{1}{\hat{\rho}_{fin}}\right)\right\}\hat{\Theta}^{\dagger} \right\rangle_R \\
&= \left\langle \left(\hat{\Theta}\hat{A}\hat{\Theta}^{\dagger}\right)\left(\hat{\Theta}\hat{U}\hat{\Theta}^{\dagger}\right)\left(\hat{\Theta}\hat{\rho}_{in}\hat{\Theta}^{\dagger}\right)\left(\hat{\Theta}\hat{U}^{\dagger}\hat{\Theta}^{\dagger}\right)\left(\hat{\Theta}\frac{1}{\hat{\rho}_{fin}}\hat{\Theta}^{\dagger}\right) \right\rangle_R \\
&= \left\langle {}^{\Theta}\hat{A}\left({}^{\Theta}\hat{U}^{\dagger\Theta}\hat{\rho}_{in}{}^{\Theta}\hat{U}\right)\frac{1}{{}^{\Theta}\hat{\rho}_{fin}} \right\rangle_R,
\end{aligned}$$

(3.10)

where we have also used (3.7), $\hat{U}\hat{U}^{\dagger} = I$ and $\hat{\Theta}^{\dagger}\hat{\Theta} = I$.

## 3.2. The Jarzynski equality from the quantum Crooks-Jarzynski relation

For $\hat{A} = I$, the quantum Crooks-Jarzynski relation, (3.6), becomes

$$\left\langle \left(\hat{U}^\dagger \hat{\rho}_{\text{fin}} \hat{U}\right) \frac{1}{\hat{\rho}_{\text{in}}} \right\rangle_F = 1 \qquad (3.11)$$

as

$$\left\langle \left(\hat{U}^\dagger \hat{\rho}_{\text{fin}} \hat{U}\right) \frac{1}{\hat{\rho}_{\text{in}}} \right\rangle_F = \langle I \rangle_R = \text{Tr}\left[{}^\Theta \hat{\rho}_{\text{fin}}\right] = \text{Tr}\left[\hat{\Theta} \hat{\rho}_{\text{fin}} \hat{\Theta}^\dagger\right] = \text{Tr}\left[\hat{\rho}_{\text{fin}}\right] = 1 ,$$

$$(3.12)$$

where we have used (3.5).

If $\hat{\rho}_{\text{fin}}$ is the canonical ensemble density matrix defined by

$$\hat{\rho}_{\text{fin}} \equiv \frac{1}{Z_{\text{fin}}(\beta)} \exp\left[-\beta \hat{H}(\tau)\right] = \exp(\beta F_{\text{fin}}) \exp\left[-\beta \hat{H}(\tau)\right], \qquad (3.13)$$

where $Z_{\text{fin}}$ defined by $Z_{\text{fin}}(\beta) \equiv \sum_f \exp\left[-\beta E(f)\right]$ is the partition function for the system and is related to the Helmholtz free energy $F_{\text{fin}}$ of the system by $Z_{\text{fin}} = \exp(-\beta F_{\text{fin}})$, then using this density matrix $\hat{\rho}_{\text{fin}}$ and $\hat{\rho}_{\text{in}}$ defined by (2.4) in (3.11), we obtain Eq. (16) in [24],

$$\left\langle e^{-\beta \hat{H}_F(\tau)} e^{\beta \hat{H}(0)} \right\rangle_F = e^{-\beta \Delta F} , \qquad (3.14)$$

where $\Delta F \equiv F_{\text{fin}} - F_{\text{in}}$. $\hat{H}_F(\tau)$ is defined by $\hat{H}_F(\tau) \equiv \hat{U}^\dagger \hat{H}(\tau) \hat{U}$, where $\hat{U}$ stands for $\hat{U}(0,\tau)$. Note that $\hat{H}(0) = \hat{H}_F(0)$ since $\hat{H}(0) = \hat{U}^\dagger(0,0) \hat{H}(0) \hat{U}(0,0) = \hat{H}_F(0)$, where $\hat{U}^\dagger(0,0) = I$.

As shown by Tasaki [26], with the forward process average defined by

$$\left\langle\!\left\langle C(i,f) \right\rangle\!\right\rangle_F \equiv \sum_{\substack{i,f \\ \rho_{\text{in}}(i) \neq 0}} C(i,f) \left|\langle f|\hat{U}|i\rangle\right|^2 \rho_{\text{in}}(i), \qquad (3.15)$$

the left-hand side of (3.14) can be rewritten as

$$\left\langle e^{-\beta \hat{H}_F(\tau)} e^{\beta \hat{H}(0)} \right\rangle_F = \left\langle\!\left\langle e^{-\beta\{E(f)-E(i)\}} \right\rangle\!\right\rangle_F \tag{3.16}$$

since

$$\begin{aligned}
&\left\langle e^{-\beta \hat{H}_F(\tau)} e^{\beta \hat{H}(0)} \right\rangle_F \\
&= \mathrm{Tr}\!\left[\hat{\rho}_{\mathrm{in}} \hat{U}^\dagger e^{-\beta \hat{H}(\tau)} \hat{U} e^{\beta \hat{H}(0)}\right] = \mathrm{Tr}\!\left[\hat{U}^\dagger e^{-\beta \hat{H}(\tau)} \hat{U} e^{\beta \hat{H}(0)} \hat{\rho}_{\mathrm{in}}\right] \\
&= \sum_{i,f} \langle i|\hat{U}^\dagger e^{-\beta \hat{H}(\tau)}|f\rangle\langle f|\hat{U} e^{\beta \hat{H}(0)} \hat{\rho}_{\mathrm{in}}|i\rangle \\
&= \sum_{i,f} e^{-\beta\{E(f)-E(i)\}} |\langle f|\hat{U}|i\rangle|^2 \rho_{\mathrm{in}}(i) = \left\langle\!\left\langle e^{-\beta\{E(f)-E(i)\}} \right\rangle\!\right\rangle_F ,
\end{aligned}$$

(3.17)

where we have used $\sum_f |f\rangle\langle f| = I$.

By defining the work $W$ done on the system during a forward process from an initial eigenstate $|i\rangle$ to a final eigenstate $|f\rangle$ by

$$W(i.f) \equiv E(f) - E(i), \tag{3.18}$$

we obtain the following Jarzynski equality from (3.14):

$$\left\langle\!\left\langle e^{-\beta W(i.f)} \right\rangle\!\right\rangle_F = e^{-\beta \Delta F}. \tag{3.19}$$

(3.11) can be also written as

$$\left\langle\!\left\langle \frac{\rho_{\mathrm{fin}}(f)}{\rho_{\mathrm{in}}(i)} \right\rangle\!\right\rangle_F = \left\langle\!\left\langle e^{-\beta\{W(i.f)-\Delta F\}} \right\rangle\!\right\rangle_F = 1, \tag{3.20}$$

where we have used (3.19) and

$$\frac{\rho_{\text{fin}}(f)}{\rho_{\text{in}}(i)} = e^{-\beta\{E(f)-E(i)\}} e^{\beta(F_{\text{fin}}-F_{\text{in}})} = e^{-\beta\{W(i.f)-\Delta F\}}. \qquad (3.21)$$

Note that (3.20) is a special case of the quantum Crooks relation, (4.4), with $C(i,f)=1$ so that the Jarzynski equality, (3.19), is also a special case of the quantum Crooks relation.

## 3.3. The standard linear response theory from the quantum Crooks-Jarzynski relation

If we assume $\hat{H}(0) = \hat{H}(\tau)$, then $\hat{\rho}_{\text{fin}} = \hat{\rho}_{\text{in}}$ and $Z_{\text{fin}} = Z_{\text{in}}$. Using (2.3) in the Crooks-Jarzynski relation, (3.6), we then obtain Eq. (18) in [24]:

$$\left\langle \hat{A}_{\text{F}}(\tau) e^{-\beta\hat{H}_{\text{F}}(\tau)} e^{\beta\hat{H}(0)} \right\rangle_{\text{F}} = \left\langle \hat{A} \right\rangle_{\text{F}} \qquad (3.22)$$

as

$$\left\langle \hat{A}_{\text{F}}(\tau) e^{-\beta\hat{H}_{\text{F}}(\tau)} e^{\beta\hat{H}(0)} \right\rangle_{\text{F}} = \left\langle {}^{\Theta}\hat{A} \right\rangle_{\text{R}} = \text{Tr}\left[ {}^{\Theta}\hat{\rho}_{\text{in}} {}^{\Theta}\hat{A} \right] = \text{Tr}\left[ \hat{\rho}_{\text{in}} \hat{A} \right] = \left\langle \hat{A} \right\rangle_{\text{F}}.$$

$$(3.23)$$

Andrieux and Gaspard [24] used (3.22) to recover the standard linear response theory with the Green-Kubo formula for a linear response coefficient for a quantity induced by the external field in terms of a canonical correlation function of the induced quantity.

We can also recover the standard linear response theory with the following special case of (3.22), where $\hat{A} = \hat{H}(\tau) = \hat{H}(0)$ and $\hat{A}_{\text{F}}(\tau) = \hat{H}_{\text{F}}(\tau)$:

$$\left\langle \hat{H}_{\text{F}}(\tau) e^{-\beta\hat{H}_{\text{F}}(\tau)} e^{\beta\hat{H}(0)} \right\rangle_{\text{F}} = \left\langle \hat{H} \right\rangle_{\text{F}}. \qquad (3.24)$$

## 3.4. The quantum Crooks-Jarzynski relation for general initial density matrices including a microcanonical ensemble density matrix

The quantum Crooks-Jarzynski relation (3.6) derived in Sec. 3.1 applies only to an initial density matrix that is positive-definite, such as the canonical density matrix defined by (2.4). In this section, we will show that for a general initial density matrix that is always *positive-semidefinite* but not necessarily *positive-definite*, such as a microcanonical ensemble density matrix, we can also derive the quantum Crooks-Jarzynski relation,

$$\left\langle \hat{A}_{C,F}(\tau)(\hat{U}^\dagger \hat{\rho}_{\text{fin}} \hat{U}) \frac{1}{\hat{\rho}_{\text{in}}} \right\rangle_F = \left\langle {}^\Theta \hat{A}_C \right\rangle_R , \qquad (3.25)$$

for any operator $\hat{A}_C$ that can be defined as

$$\begin{aligned}\hat{A}_C &\equiv \sum_{i,f,f'} |f'\rangle\langle f'|\hat{U}|i\rangle C(i,f)\langle i|\hat{U}^\dagger|f\rangle\langle f| \\ &= \sum_{i,f} \hat{U}|i\rangle C(i,f)\langle i|\hat{U}^\dagger|f\rangle\langle f| ,\end{aligned}$$

$$(3.26)$$

where $C(i,f)$ is a complex-valued quantity that vanishes or $C(i,f)=0$ for each initial eigenstate $|i\rangle$ for which the initial statistical distribution vanishes or $\rho_{\text{in}}(i)=0$.

Using

$$\hat{A}_{C,F}(\tau) = \hat{U}^\dagger \hat{A}_C \hat{U} = \sum_{i,f} |i\rangle C(i,f) \langle i|\hat{U}^\dagger |f\rangle \langle f|\hat{U}, \qquad (3.27)$$

we can show the quantum Crooks-Jarzynski relation (3.25) for $\hat{A}_C$ as follows:

$$\begin{aligned}
&\left\langle \hat{A}_{C,F}(\tau)\left(\hat{U}^\dagger \hat{\rho}_{\text{fin}} \hat{U}\right)\frac{1}{\hat{\rho}_{\text{in}}}\right\rangle_F \\
&= \sum_{\substack{i \\ \rho_{\text{in}}(i)\neq 0}} \rho_{\text{in}}(i)\langle i|\hat{A}_{C,F}(\tau)\left(\hat{U}^\dagger \hat{\rho}_{\text{fin}} \hat{U}\right)\frac{1}{\hat{\rho}_{\text{in}}}|i\rangle \\
&= \sum_{\substack{i \\ \rho_{\text{in}}(i)\neq 0}} \rho_{\text{in}}(i)\langle i|\left\{\sum_{i',f}|i'\rangle C(i',f)\langle i'|\hat{U}^\dagger |f\rangle \langle f|\hat{U}\right\}\left(\hat{U}^\dagger \hat{\rho}_{\text{fin}} \hat{U}\right)|i\rangle \frac{1}{\rho_{\text{in}}(i)} \\
&= \sum_{\substack{i,f \\ \rho_{\text{in}}(i)\neq 0}} C(i,f)\langle i|\hat{U}^\dagger |f\rangle \langle f|\hat{U}|i\rangle \rho_{\text{fin}}(f) \\
&= \sum_{i,f} C(i,f)\langle i|\hat{U}^\dagger |f\rangle \langle f|\hat{U}|i\rangle \rho_{\text{fin}}(f) \\
&= \sum_f \langle f|\left\{\sum_{i,f'} \hat{U}|i\rangle C(i,f')\langle i|\hat{U}^\dagger |f'\rangle \langle f'|\right\}|f\rangle \rho_{\text{fin}}(f) \\
&= \sum_f \langle f|\hat{A}_C|f\rangle \rho_{\text{fin}}(f) = \sum_f \langle f|\hat{A}_C \hat{\rho}_{\text{fin}}|f\rangle \\
&= \text{Tr}\left[\hat{A}_C \hat{\rho}_{\text{fin}}\right] = \text{Tr}\left[\hat{\rho}_{\text{fin}} \hat{A}_C\right] \\
&= \text{Tr}\left[\hat{\Theta}\hat{\rho}_{\text{fin}} \hat{\Theta}^\dagger \hat{\Theta}\hat{A}\hat{\Theta}^\dagger\right] = \text{Tr}\left[{}^\Theta \hat{\rho}_{\text{fin}} {}^\Theta \hat{A}\right] = \left\langle {}^\Theta \hat{A}\right\rangle_R,
\end{aligned}$$

$$(3.28)$$

where we have used $\hat{U}\hat{U}^\dagger = I$, $\langle i|i'\rangle = \delta_{i,i'}$, $\langle f'|f\rangle = \delta_{f',f}$, and $\hat{\Theta}^\dagger \hat{\Theta} = I$. We have also used the condition for $C(i,f)$ that it vanishes or $C(i,f)=0$ if $\rho_{\text{in}}(i)=0$.

In Sec. 4.1, we will derive the quantum Crooks relation for the quantity $C(i,f)$ and then in Sec. 4.2, we will show that it is equivalent to the quantum

Crooks-Jarzynski relation (3.25) for the corresponding operator $\hat{A}_C$ so that using (3.25), we can derive any result that follows from the quantum Crooks relation with $C(i,f)$ and vice versa.

Note that the operator $\hat{A}_{C,F}(\tau)$ satisfies

$$\left\langle \hat{A}_{C,F}(\tau) \right\rangle_F = \left\langle\!\left\langle C(i,f) \right\rangle\!\right\rangle_F \qquad (3.29)$$

as

$$\begin{aligned}
\left\langle \hat{A}_{C,F}(\tau) \right\rangle_F &= \sum_{\substack{i \\ \rho_{in}(i)\neq 0}} \rho_{in}(i) \langle i | \hat{A}_{C,F}(\tau) | i \rangle \\
&= \sum_{\substack{i \\ \rho_{in}(i)\neq 0}} \rho_{in}(i) \langle i | \left\{ \sum_{i',f} |i'\rangle C(i',f) \langle i' | \hat{U}^\dagger | f \rangle \langle f | \hat{U} \right\} | i \rangle \\
&= \sum_{\substack{i \\ \rho_{in}(i)\neq 0}} C(i,f) \left| \langle f | \hat{U} | i \rangle \right|^2 \rho_{in}(i) \\
&= \left\langle\!\left\langle C(i,f) \right\rangle\!\right\rangle_F ,
\end{aligned}$$

(3.30)

where we have used $\langle i | i' \rangle = \delta_{i,i'}$.

Not every operator can be expressed in terms of $C(i,f)$ as in (3.26). If an operator $\hat{A}_C$ is defined by (3.26), then it must satisfy

$$\langle i | \hat{U}^\dagger \hat{A}_C | f \rangle = C(i,f) \langle i | \hat{U}^\dagger | f \rangle \qquad (3.31)$$

since

$$\langle i|\hat{U}^{\dagger}\hat{A}_{C}|f\rangle = \langle i|\hat{U}^{\dagger}\left\{\sum_{i',f'}\hat{U}|i'\rangle C(i',f')\langle i'|\hat{U}^{\dagger}|f'\rangle\langle f'|\right\}|f\rangle$$
$$= \sum_{i'}\langle i|\hat{U}^{\dagger}\hat{U}|i'\rangle C(i',f)\langle i'|\hat{U}^{\dagger}|f\rangle$$
$$= C(i,f)\langle i|\hat{U}^{\dagger}|f\rangle,$$

(3.32)

where we have used $\hat{U}^{\dagger}\hat{U}=I$, $\langle i|i'\rangle = \delta_{i,i'}$, and $\langle f'|f\rangle = \delta_{f',f}$. (3.31) implies that $\langle i|\hat{U}^{\dagger}\hat{A}_{C}|f\rangle = 0$ if $\langle i|\hat{U}^{\dagger}|f\rangle = 0$, which does not generally hold for any operator.

According to (3.31), the operator $\hat{A}_{C}$ is uniquely determined by the values of $C(i,f)$ except for its values for pairs, if any, of initial and final eigenstates that satisfy $\langle i|\hat{U}^{\dagger}|f\rangle = 0$. Suppose that for two quantities $C(i,f)$ and $C'(i,f)$, we find $\hat{A}_{C} = \hat{A}_{C'}$. Using (3.31), we then obtain

$$\{C(i,f)-C'(i,f)\}\langle i|\hat{U}^{\dagger}|f\rangle = \langle i|\hat{U}^{\dagger}\hat{A}_{C}|f\rangle - \langle i|\hat{U}^{\dagger}\hat{A}_{C'}|f\rangle = 0,$$

(3.33)

which implies that $C(i,f) = C'(i,f)$ except when $\langle i|\hat{U}^{\dagger}|f\rangle = 0$.

Note that the definition of the operator $\hat{A}_{C}$ in terms of $C(i,f)$ is not as restrictive as it may seem. If the initial statistical distribution is the canonical ensemble distribution defined by (2.4), which is positive-definite so that $\rho_{\text{in}}(i) \neq 0$ for every initial eigenstate, $C(i,f)$ can be in fact any arbitrary complex-valued quantity.

For $C(i,f)=1$, the corresponding operator is $\hat{A}_{1}=I$ as

$$\hat{A}_I = \sum_{i,f} \hat{U}|i\rangle\langle i|\hat{U}^\dagger|f\rangle\langle f| = I, \qquad (3.34)$$

where we have used $\sum_i |i\rangle\langle i| = I$, $\sum_f |f\rangle\langle f| = I$, and $\hat{U}\hat{U}^\dagger = I$. The quantum Crooks-Jarzynski relation (3.25) with this operator is then nothing but (3.11) that leads to the Jarzynski equality (3.19).

As another example, consider $C(i,f) = E(f)$. The corresponding operator is then $\hat{A}_{E(f)} = \hat{H}(\tau)$ as

$$\hat{A}_{E(f)} = \sum_{i,f} \hat{U}|i\rangle E(f)\langle i|\hat{U}^\dagger|f\rangle\langle f| = \sum_{i,f} \hat{U}|i\rangle\langle i|\hat{U}^\dagger \hat{H}(\tau)|f\rangle\langle f| = \hat{H}(\tau).$$

(3.35)

If we assume $\hat{H}(\tau) = \hat{H}(0)$, then the quantum Crooks-Jarzynski relation (3.25) with this operator is nothing but (3.24) that leads to the standard linear response theory.

Suppose the initial density matrix is the following microcanonical ensemble density matrix defined by

$$\hat{\rho}_{\text{in}}^{\text{mic}} \equiv \sum_i |i\rangle \frac{\delta_\Delta(E(i) - E_0)}{\exp(S_0/k_B)} \langle i|. \qquad (3.36)$$

The "regularized delta-function" $\delta_\Delta(E(i) - E_0)$ is defined by

$$\delta_\Delta(E(i) - E_0) \equiv \begin{cases} 1 & (\text{if } E_0 \leq E(i) \leq E_0 + \Delta) \\ 0 & (\text{otherwise}) \end{cases}, \quad (3.37)$$

where we assume $\Delta \ll |E_0|$. $S_0$ is the Boltzmann entropy for the system at an initial internal energy $E_0$ and satisfies

$$\exp(S_0/k_B) = \sum_i \delta_\Delta(E(i) - E_0). \quad (3.38)$$

The quantum Crooks-Jarzynski relation (3.25) then holds for any operator $\hat{A}_C$ corresponding to $C(i,f)$ defined by

$$C(i,f) \equiv \delta_\Delta(E(i) - E_0)\tilde{C}(i,f), \quad (3.39)$$

where $\tilde{C}(i,f)$ is an arbitrary complex-valued quantity.

In Sec. 4.3, using quantum Crooks relation (4.4) for $C(i,f) \equiv \delta_\Delta(E(i) - E_0)$, we will derive the Crooks transient fluctuation theorem for a system whose initial statistical distribution is the following microcanonical ensemble distribution [31]:

$$\rho_{\text{in}}^{\text{mic}}(i) = \frac{\delta_\Delta(E(i) - E_0)}{\exp(S_0/k_B)}, \quad (3.40)$$

which satisfies $\hat{\rho}_{in}^{mic}|i\rangle = \rho_{in}^{mic}(i)|i\rangle$.

# 4. THE QUANTUM CROOKS RELATION

## 4.1. Derivation of the quantum Crooks relation

In this section, we will derive the quantum extension of the Crooks relation, which relates the forward process average of a product of quantities including a complex-valued quantity $C(i,f)$, which may depend on the initial and the final eigenstates of a system and vanishes for an initial eigenstate for which $\rho_{in}$ vanishes, to the backward process average of the same quantity $C(i,f)$.

The forward process average $\langle\langle\ \rangle\rangle_F$ has been defined by (3.15) while the backward process average $\langle\langle\ \rangle\rangle_R$ is defined by

$$\langle\langle C(i,f)\rangle\rangle_R \equiv \sum_{\Theta_i,\Theta_f} C(i,f)\left|\langle\Theta i|\Theta\hat{U}|\Theta f\rangle\right|^2 \Theta\rho_{fin}(\Theta f). \qquad (4.1)$$

$\rho_{fin}(f)$ is the statistical distribution corresponding to a density matrix $\hat{\rho}_{fin}$ for the final eigenstates of the system and we assume that $\hat{\rho}_{fin}$ and $\rho_{fin}(f)$ satisfy

$$\hat{\rho}_{fin}|f\rangle = \rho_{fin}(f)|f\rangle. \qquad (4.2)$$

According to (2.16), $\Theta\rho_{fin}(\Theta f)$ defined by $\Theta\hat{\rho}_{fin}|\Theta f\rangle = \Theta\rho_{fin}(\Theta f)|\Theta f\rangle$ then satisfies

$$\Theta\rho_{fin}(\Theta f) = \rho_{fin}(f). \qquad (4.3)$$

The quantum Crooks relation is then the following general identity for $C(i,f)$ that vanishes or $C(i,f)=0$ if $\rho_{in}(i)=0$:

$$\left\langle\!\!\left\langle C(i,f)\frac{\rho_{fin}(f)}{\rho_{in}(i)}\right\rangle\!\!\right\rangle_F = \left\langle\!\!\left\langle C(i,f)\right\rangle\!\!\right\rangle_R, \tag{4.4}$$

where $\rho_{in}$ is the statistical distribution corresponding to the initial density matrix $\hat{\rho}_{in}$. For the canonical ensemble density matrix defined by (2.4), $\rho_{in}$ is always positive or $\rho_{in}(i)>0$ for any initial eigenstate so that the quantum Crooks relation holds for any complex-valued quantity $C(i,f)$.

We can show this quantum Crooks relation as follows.

$$\begin{aligned}\left\langle\!\!\left\langle C(i,f)\frac{\rho_{fin}(f)}{\rho_{in}(i)}\right\rangle\!\!\right\rangle_F &= \sum_{\substack{i,f \\ \rho_{in}(i)\neq 0}} C(i,f)\frac{\rho_{fin}(f)}{\rho_{in}(i)}\left|\langle f|\hat{U}|i\rangle\right|^2 \rho_{in}(i) \\ &= \sum_{\substack{i,f \\ \rho_{in}(i)\neq 0}} C(i,f)\left|\langle {}^\Theta i|{}^\Theta \hat{U}|{}^\Theta f\rangle\right| \rho_{fin}(f) \\ &= \sum_{i,f} C(i,f)\left|\langle {}^\Theta i|{}^\Theta \hat{U}|{}^\Theta f\rangle\right| \rho_{fin}(f) \\ &= \sum_{{}^\Theta i,{}^\Theta f} C(i,f)\left|\langle {}^\Theta i|{}^\Theta \hat{U}|{}^\Theta f\rangle\right|^2 \rho_{fin}(f) \\ &= \sum_{{}^\Theta i,{}^\Theta f} C(i,f)\left|\langle {}^\Theta i|{}^\Theta \hat{U}|{}^\Theta f\rangle\right|^2 {}^\Theta\!\rho_{fin}({}^\Theta f) \\ &= \left\langle\!\!\left\langle C(i,f)\right\rangle\!\!\right\rangle_R, \end{aligned} \tag{4.5}$$

where we have used (4.3), and (2.13), the direct consequence of the principle of microreversibility. We have also used the condition for $C(i,f)$ that it vanishes or $C(i,f)=0$ if $\rho_{in}(i)=0$. Note that the time reversal operator $\hat{\Theta}$ provides a one-to-one map from the set of all the initial eigenstates of $\hat{H}(0)$ to itself as well as from the set of all the final eigenstates of $\hat{H}(\tau)$ to itself and that $\{|^{\Theta}i\rangle\} = \{|i\rangle\}$ and $\{|^{\Theta}f\rangle\} = \{|f\rangle\}$.

As this derivation shows, the quantum Crooks relation is simply based on (2.13), $|\langle f|\hat{U}|i\rangle|^2 = |\langle ^{\Theta}i|^{\Theta}\hat{U}|^{\Theta}f\rangle|^2$, and (4.3), $\rho_{\text{fin}}(f) = {}^{\Theta}\rho_{\text{fin}}({}^{\Theta}f)$. It is rather remarkable that such a deceptively simple relation leads to highly nontrivial results such as the Crooks transient fluctuation theorem (see Sec. 4.3) and the fluctuation theorem for the shear stress on a fluid in a steady shear flow, which leads to the Green-Kubo formula for the shear viscosity of the fluid (see Sec. 4.4).

Note also that the quantum Crooks relation is a general mathematical identity that holds for arbitrary statistical distributions $\rho_{\text{in}}$ and $\rho_{\text{fin}}$ and for any arbitrary quantity $C(i,f)$ that vanishes for an initial eigenstate for which $\rho_{\text{in}}$ vanishes. Furthermore, the quantity $C(i,f)$ does not need to be a physically observable quantity.

In addition, the statistical distribution $\rho_{\text{in}}$ or $\rho_{\text{fin}}$ does not need to represent an equilibrium ensemble for the initial or the final eigenstates for the system, which suggests that we may extend the quantum Crooks relation to non-equilibrium ensembles that are represented by some statistical distributions.

**4.2. Quantum Crooks-Jarzynski relation equivalent to a quantum Crooks relation**

In Sec. 3.4, we have shown that for a complex-valued quantity $C(i,f)$ that vanishes or $C(i,f) = 0$ for each initial eigenstate $|i\rangle$ for which the initial statistical distribution vanishes or $\rho_{\text{in}}(i) = 0$, we can define a corresponding operator $\hat{A}_C$ by (3.26) and that the operator $\hat{A}_C$ satisfies a quantum Crooks-Jarzynski relation (3.25). In Sec. 4.1, we have also shown that the quantity $C(i,f)$ satisfies a quantum Crooks relation (4.4).

In this section, we will show that within the framework of quantum dynamics, where the principle of microreversibility always holds, the quantum Crooks relation for $C(i,f)$ and the corresponding quantum Crooks-Jarzynski relation for $\hat{A}_C$ are equivalent to each other so that using the Crooks-Jarzynski relation for $\hat{A}_C$, we can derive any result that follows from the quantum Crooks relation for $C(i,f)$ and vice versa.

First, for $C(i,f)$ and $\hat{A}_C$ defined by (3.26), we can show

$$\left\langle \hat{A}_{C,\text{F}}(\tau)\left(\hat{U}^{\dagger}\hat{\rho}_{\text{fin}}\hat{U}\right)\frac{1}{\hat{\rho}_{\text{in}}}\right\rangle_{\text{F}} = \left\langle\!\!\left\langle C(i,f)\frac{\rho_{\text{fin}}(f)}{\rho_{\text{in}}(i)}\right\rangle\!\!\right\rangle_{\text{F}} \tag{4.6}$$

since

$$\begin{aligned}
&\left\langle \hat{A}_{C,\text{F}}(\tau)\left(\hat{U}^{\dagger}\hat{\rho}_{\text{fin}}\hat{U}\right)\frac{1}{\hat{\rho}_{\text{in}}}\right\rangle_{\text{F}} \\
&= \sum_{\substack{i \\ \rho_{\text{in}}(i)\neq 0}} \rho_{\text{in}}(i)\langle i|\hat{A}_{C,\text{F}}(\tau)\left(\hat{U}^{\dagger}\hat{\rho}_{\text{fin}}\hat{U}\right)\frac{1}{\hat{\rho}_{\text{in}}}|i\rangle \\
&= \sum_{\substack{i \\ \rho_{\text{in}}(i)\neq 0}} \rho_{\text{in}}(i)\langle i|\left\{\sum_{i',f}|i'\rangle C(i',f)\langle i'|\hat{U}^{\dagger}|f\rangle\langle f|\hat{U}\right\}\left(\hat{U}^{\dagger}\hat{\rho}_{\text{fin}}\hat{U}\right)|i\rangle\frac{1}{\rho_{\text{in}}(i)} \\
&= \sum_{\substack{i,i',f \\ \rho_{\text{in}}(i)\neq 0}} \delta_{i,i'}C(i',f)\langle i'|\hat{U}^{\dagger}|f\rangle\langle f|\hat{\rho}_{\text{fin}}\hat{U}|i\rangle\frac{1}{\rho_{\text{in}}(i)}\rho_{\text{in}}(i) \\
&= \sum_{\substack{i,f \\ \rho_{\text{in}}(i)\neq 0}} C(i,f)\langle i|\hat{U}^{\dagger}|f\rangle\langle f|\hat{U}|i\rangle\frac{\rho_{\text{fin}}(f)}{\rho_{\text{in}}(i)}\rho_{\text{in}}(i) \\
&= \sum_{\substack{i,f \\ \rho_{\text{in}}(i)\neq 0}} C(i,f)|\langle f|\hat{U}|i\rangle|^2\frac{\rho_{\text{fin}}(f)}{\rho_{\text{in}}(i)}\rho_{\text{in}}(i) \\
&= \left\langle\!\!\left\langle C(i,f)\frac{\rho_{\text{fin}}(f)}{\rho_{\text{in}}(i)}\right\rangle\!\!\right\rangle_{\text{F}},
\end{aligned}$$

$$\tag{4.7}$$

where we have also used (3.27), (4.2), $\sum_f |f\rangle\langle f| = I$, and $\hat{U}\hat{U}^\dagger = I$.

Second, for $C(i,f)$ and $\hat{A}_C$, we also find

$$\langle {}^\Theta \hat{A}_C \rangle_R = \langle\langle C(i,f) \rangle\rangle_R \qquad (4.8)$$

where

$$\begin{aligned}
{}^\Theta \hat{A}_C &= \hat{\Theta} \hat{A}_C \hat{\Theta}^\dagger \\
&= \hat{\Theta} \left\{ \sum_{i,f,f'} |f'\rangle\langle f'|\hat{U}|i\rangle C(i,f) \langle i|\hat{U}^\dagger|f\rangle\langle f| \right\} \hat{\Theta}^\dagger \\
&= \sum_{i,f,f'} |{}^\Theta f'\rangle\langle {}^\Theta i|{}^\Theta \hat{U}|{}^\Theta f'\rangle C(i,f) \langle {}^\Theta f|{}^\Theta \hat{U}^\dagger|{}^\Theta i\rangle\langle f|\hat{\Theta}^\dagger ,
\end{aligned}$$

(4.9)

where we have used (2.14), a consequence of the principle of microreversibility.

We can show (4.8) as follows.

$$\begin{aligned}
&\langle {}^\Theta \hat{A}_C \rangle_R \\
&= \text{Tr}[{}^\Theta \hat{\rho}_{\text{fin}} {}^\Theta \hat{A}_C] = \text{Tr}[{}^\Theta \hat{A}_C {}^\Theta \hat{\rho}_{\text{fin}}] \\
&= \sum_{{}^\Theta f} \langle {}^\Theta f|{}^\Theta \hat{A}_C {}^\Theta \hat{\rho}_{\text{fin}}|{}^\Theta f\rangle = \sum_{{}^\Theta f} \langle {}^\Theta f|{}^\Theta \hat{A}_C|{}^\Theta f\rangle {}^\Theta \rho_{\text{fin}}({}^\Theta f) \\
&= \sum_{{}^\Theta f} \langle {}^\Theta f| \left\{ \sum_{{}^\Theta i, {}^\Theta f', {}^\Theta f''} |{}^\Theta f'\rangle\langle {}^\Theta i|{}^\Theta \hat{U}|{}^\Theta f'\rangle C(i,f'') \langle {}^\Theta f''|{}^\Theta \hat{U}^\dagger|{}^\Theta i\rangle\langle f''|\hat{\Theta}^\dagger \right\} |{}^\Theta f\rangle {}^\Theta \rho_{\text{fin}}({}^\Theta f) \\
&= \sum_{{}^\Theta i, {}^\Theta f} C(i,f) |\langle {}^\Theta i|{}^\Theta \hat{U}|{}^\Theta f\rangle|^2 {}^\Theta \rho_{\text{fin}}({}^\Theta f) \\
&= \langle\langle C(i,f) \rangle\rangle_R ,
\end{aligned}$$

(4.10)

where we have used $\langle {}^\Theta f|{}^\Theta f'\rangle = \delta_{{}^\Theta f, {}^\Theta f'}$ and $\langle f''|f\rangle = \delta_{f'',f}$.

If we assume that the quantum Crooks relation (4.4) holds for any arbitrary quantity $C(i,f)$, then for its corresponding operator $\hat{A}_C$, we can show that the quantum Crooks-Jarzynski relation (3.26) holds as

$$\left\langle \hat{A}_{C,\mathrm{F}}(\tau)(\hat{U}^\dagger \hat{\rho}_{\mathrm{fin}} \hat{U}) \frac{1}{\hat{\rho}_{\mathrm{in}}} \right\rangle_{\mathrm{F}} = \left\langle\!\!\left\langle C(i,f) \frac{\rho_{\mathrm{fin}}(f)}{\rho_{\mathrm{in}}(i)} \right\rangle\!\!\right\rangle_{\mathrm{F}} = \left\langle\!\!\left\langle C(i,f) \right\rangle\!\!\right\rangle_{\mathrm{R}} = \left\langle {}^\Theta \hat{A}_C \right\rangle_{\mathrm{R}}.$$

(4.11)

On the other hand, if we assume that the quantum Crooks-Jarzynski relation (3.26) holds for operator $\hat{A}_C$, then for $C(i,f)$, we can show that the quantum Crooks relation (4.4) holds as

$$\left\langle\!\!\left\langle C(i,f) \frac{\rho_{\mathrm{fin}}(f)}{\rho_{\mathrm{in}}(i)} \right\rangle\!\!\right\rangle_{\mathrm{F}} = \left\langle \hat{A}_{C,\mathrm{F}}(\tau)(\hat{U}^\dagger \hat{\rho}_{\mathrm{fin}} \hat{U}) \frac{1}{\hat{\rho}_{\mathrm{in}}} \right\rangle_{\mathrm{F}} = \left\langle {}^\Theta \hat{A}_C \right\rangle_{\mathrm{R}} = \left\langle\!\!\left\langle C(i,f) \right\rangle\!\!\right\rangle_{\mathrm{R}}.$$

(4.12)

We have thus shown that the quantum Crooks relation (4.4) for $C(i,f)$ and the corresponding quantum Crooks-Jarzynski relation (3.26) for $\hat{A}_C$ are equivalent to each other.

As mentioned in Sec. 3.4, the operator corresponding to $C(i,f) = E(f)$ is $\hat{A}_{E(f)} = \hat{H}(\tau)$ and the quantum Crooks-Jarzynski relation for this operator is nothing but (3.24) that leads to the standard linear response theory. Being equivalent to this quantum Crooks-Jarzynski relation for $\hat{A}_{E(f)} = \hat{H}(\tau)$, the quantum Crooks relation for $C(i,f) = E(f)$ also leads to the standard linear response theory.

### 4.3. The Crooks transient fluctuation theorem from the quantum Crooks relation

For classical systems, Crooks [16] showed that we can derive the Crooks transient fluctuation theorem from the Crooks relation. In this section, by extending his derivation to quantum systems, we will derive the Crooks transient fluctuation theorem for quantum systems using the quantum Crooks relation (4.4).

The Crooks transient fluctuation theorem relates the probability $P_{\text{F}}(\overline{W})$ for $W(i,f)$, which is defined by (3.18) as the work done on the system during a forward process from an initial eigenstate $|i\rangle$ to a final eigenstate $|f\rangle$, to take a specific value of $\overline{W}$ to the probability $P_{\text{R}}(-\overline{W})$ for the work $W(^{\Theta}f,^{\Theta}i)$ done on the system during a time-reversed backward process from an initial eigenstate $|^{\Theta}f\rangle$ to a final eigenstate $|^{\Theta}i\rangle$ to take a value of $-\overline{W}$. According to (2.16), $|^{\Theta}i\rangle$ is an eigenstate of $^{\Theta}\hat{H}(0)$ with energy eigenvalue $E(i)$ and $|^{\Theta}f\rangle$ is an eigenstate of $^{\Theta}\hat{H}(\tau)$ with energy eigenvalue $E(f)$ so that

$$W(^{\Theta}f,^{\Theta}i) = E(i) - E(f) = -W(i,f). \qquad (4.13)$$

The probability $P_{\text{F}}(\overline{W})$ is defined by

$$P_{\text{F}}(\overline{W}) \equiv \sum_{i,f} \delta(W(i,f) - \overline{W}) |\langle f|\hat{U}|i\rangle|^2 \rho_{\text{in}}(i) = \langle\langle \delta(W(i,f) - \overline{W}) \rangle\rangle_{\text{F}}$$

$$(4.14)$$

while the probability $P_{\text{R}}(-\overline{W})$ is defined by

$$P_R(-\overline{W}) \equiv \sum_{\Theta_i, \Theta_f} \delta(W(^\Theta f, ^\Theta i) + \overline{W}) |\langle ^\Theta i|^\Theta \hat{U}|^\Theta f\rangle|^2 {}^\Theta\rho_{\text{fin}}(^\Theta f)$$

$$= \sum_{\Theta_i, \Theta_f} \delta(-W(i,f) + \overline{W}) |\langle ^\Theta i|^\Theta \hat{U}|^\Theta f\rangle|^2 {}^\Theta\rho_{\text{fin}}(^\Theta f)$$

$$= \langle\langle \delta(W(i,f) - \overline{W})\rangle\rangle_R .$$

(4.15)

Using $C(i,f) = \delta(W(i,f) - \overline{W})$ in the quantum Crooks relation, (4.4), we can then show the following Crooks transient fluctuation theorem:

$$\frac{P_R(-\overline{W})}{P_F(\overline{W})} = e^{-\beta(\overline{W} - \Delta F)}. \qquad (4.16)$$

since

$$P_R(-\overline{W}) = \langle\langle \delta(W(i,f) - \overline{W})\rangle\rangle_R = \left\langle\left\langle \delta(W(i,f) - \overline{W}) \frac{\rho_{\text{fin}}(f)}{\rho_{\text{in}}(i)}\right\rangle\right\rangle_F$$

$$= \langle\langle \delta(W(i,f) - \overline{W}) e^{-\beta\{W(i,f) - \Delta F\}}\rangle\rangle_F$$

$$= e^{-\beta(\overline{W} - \Delta F)} \langle\langle \delta(W(i,f) - \overline{W})\rangle\rangle_F$$

$$= e^{-\beta(\overline{W} - \Delta F)} P_F(\overline{W}) ,$$

(4.17)

where we have also used (3.21).

Using the quantum Crooks relation, we can also derive the Crooks transient fluctuation theorem when the initial statistical distribution is the microcanonical ensemble distribution defined by (3.40) [31] and the final statistical distribution is the following microcanonical ensemble distribution defined by

$$\rho_{\text{fin}}^{\text{mic}}(f) \equiv \frac{\delta_\Delta(E(f) - E_0 - \overline{W})}{\exp(S_\tau / k_B)}. \qquad (4.18)$$

The "regularized delta-function" $\delta_\Delta(E(f) - E_0 - \overline{W})$ is defined by

$$\delta_\Delta(E(f) - E_0 - \overline{W}) \equiv \begin{cases} 1 & (\text{if } E_0 + \overline{W} \leq E(i) \leq E_0 + \overline{W} + \Delta) \\ 0 & (\text{otherwise}) \end{cases},$$

$$(4.19)$$

where we assume $\Delta \ll |E_0 + \overline{W}|$. $S_\tau$ is the Boltzmann entropy for the system at an internal energy $E_0 + \overline{W}$ and satisfies

$$\exp(S_\tau / k_B) = \sum_f \delta_\Delta(E(f) - E_0 - \overline{W}). \qquad (4.20)$$

The Crooks transient fluctuation theorem for a system starting with the microcanonical ensemble distribution (3.40) then relates the probability $P_F(E_0, E_0 + \overline{W})$ for the system to receive a specific amount of work $\overline{W}$ during a forward process to the probability $P_R(E_0 + \overline{W}, E_0)$ for the system to receive work $-\overline{W}$ during a time-reversed backward process, where the system starts with the microcanonical ensemble distribution $^\Theta\rho_{\text{fin}}^{\text{mic}}(^\Theta f) = \rho_{\text{fin}}^{\text{mic}}(f)$.

The probability $P_F(E_0, E_0 + \overline{W})$ is defined by

$$P_{\text{F}}(E_0, E_0 + \overline{W}) \equiv \sum_{i,f} \delta_\Delta(E(f) - E_0 - \overline{W}) |\langle f|\hat{U}|i\rangle|^2 \rho_{\text{in}}^{\text{mic}}(i)$$

$$= \sum_{i,f} \delta_\Delta(E(f) - E_0 - \overline{W}) |\langle f|\hat{U}|i\rangle|^2 \frac{\delta_\Delta(E(i) - E_0)}{\exp(S_0/k_{\text{B}})}$$

$$= \langle\langle \delta_\Delta(E(f) - E_0 - \overline{W}) \rangle\rangle_{\text{F}}$$

(4.21)

while the probability $P_{\text{R}}(E_0 + \overline{W}, E_0)$ is defined by

$$P_{\text{R}}(E_0 + \overline{W}, E_0)$$
$$\equiv \sum_{\Theta i, \Theta f} \delta_\Delta(E(i) - E_0) |\langle \Theta i|^\Theta \hat{U}|^\Theta f\rangle|^2 {}^\Theta\rho_{\text{fin}}^{\text{mic}}({}^\Theta f)$$
$$= \sum_{\Theta i, \Theta f} \delta_\Delta(E(i) - E_0) |\langle \Theta i|^\Theta \hat{U}|^\Theta f\rangle|^2 \frac{\delta_\Delta(E(f) - E_0 - \overline{W})}{\exp(S_\tau/k_{\text{B}})}$$
$$= \langle\langle \delta_\Delta(E(i) - E_0) \rangle\rangle_{\text{R}} .$$

(4.22)

Using $C(i,f) \equiv \delta_\Delta(E(i) - E_0)$ in the quantum Crooks relation (4.4), we can then show the following Crooks transient fluctuation theorem:

$$\frac{P_{\text{R}}(E_0 + \overline{W}, E_0)}{P_{\text{F}}(E_0, E_0 + \overline{W})} = e^{-(S_\tau - S_0)/k_{\text{B}}} \qquad (4.23)$$

since

$$P_R(E_0 + \overline{W}, E_0) = \langle\langle \delta_\Delta(E(i) - E_0)\rangle\rangle_R$$

$$= \left\langle\left\langle \delta_\Delta(E(i) - E_0)\frac{\rho_{\text{fin}}^{\text{mic}}(f)}{\rho_{\text{in}}^{\text{mic}}(i)}\right\rangle\right\rangle_F$$

$$= \langle\langle \delta_\Delta(E(f) - E_0 - \overline{W})e^{-(S_\tau - S_0)/k_B}\rangle\rangle_F$$

$$= e^{-(S_\tau - S_0)/k_B}\langle\langle \delta_\Delta(E(f) - E_0 - \overline{W})\rangle\rangle_F$$

$$= e^{-(S_\tau - S_0)/k_B} P_F(E_0, E_0 + \overline{W}) \ .$$

(4.24)

## 4.4. The fluctuation theorem for shear stress from the quantum Crooks relation

If we assume $\hat{H}(0) = \hat{H}(\tau)$, then $\hat{\rho}_{\text{fin}} = \hat{\rho}_{\text{in}}$ so that $\Delta F = F_{\text{fin}} - F_{\text{in}} = 0$. The quantum Crooks relation, (4.4), then becomes

$$\langle\langle C(i,f)e^{-\beta W(i,f)}\rangle\rangle_F = \langle\langle C(i,f)\rangle\rangle_R, \qquad (4.25)$$

where we have used (3.21).

For a fluid in a steady shear flow driven by a constant velocity of a solid plate moving above the fluid, using this special case of the quantum Crooks relation, we can derive [30] the fluctuation theorem for the shear stress on the fluid and obtain the Green-Kubo formula for its shear viscosity $\eta$ in terms of the symmetrized correlation function of its shear stress operator $\tilde{P}_F$ in the Heisenberg picture:

$$\eta = \frac{V}{k_B T} \lim_{\tau \to \infty} \frac{1}{\tau}\int_0^\tau dt_1 \int_0^{t_1} dt_2 \left\langle \frac{1}{2}\{\tilde{P}_F(t_1)\tilde{P}_F(t_2) + \tilde{P}_F(t_2)\tilde{P}_F(t_1)\}\right\rangle_{\text{eq}}, \qquad (4.26)$$

where *V* is the volume of the fluid and the subscript "eq" indicates that the statistical average is taken when the plate remains at rest so that the fluid, the plate, and a heat reservoir attached to the fluid are all in equilibrium at the same temperature *T*.

## 5. CONCLUSIONS

In this article, we have derived a quantum extension of the Crooks-Jarzynski relation without explicitly using the principle of microreversibility, which is consistent with the fact that the principle of microreversibility was not used explicitly in the original derivations of the standard linear response theory and the Jarzynski equality.

For quantum systems driven out of equilibrium, we can derive the standard linear response theory and the Jarzynski equality directly from the quantum Crooks-Jarzynski relation without explicitly using the principle of microreversibility while we need to use the quantum Crooks relation based on the principle of microreversibility to derive the Crooks transient fluctuation theorem and the fluctuation theorem for the current or shear stress.

We have shown that for each quantity $C(i,f)$ that satisfies the quantum Crooks relation, we can define a corresponding operator $\hat{A}_C$ that satisfies a quantum Crooks-Jarzynski relation. We have also shown that the quantum Crooks relation for $C(i,f)$ and quantum Crooks-Jarzynski relation for $\hat{A}_C$ are equivalent to each other so that using the quantum Crooks-Jarzynski relation for $\hat{A}_C$, we can derive any result that follows from the quantum Crooks relation for $C(i,f)$ and vice versa.

Therefore, using either the quantum Crooks-Jarzynski relation or the quantum Crooks relation, we can derive the fluctuation relations mentioned above and the standard linear response theory.

Both the quantum Crooks relation and the quantum Crooks-Jarzynski relation are quite general mathematical identities and neither the density matrices $\hat{\rho}_{in}$ and $\hat{\rho}_{fin}$ involved in the quantum Crooks-Jarzynski relation nor the statistical distributions $\rho_{in}$ and $\rho_{fin}$ involved in the quantum Crooks relation need to represent equilibrium ensembles. This suggests that we may be able to extend these relations to quantum systems evolving from a non-equilibrium state to another. Whether such an extension of the quantum Crooks relation or the quantum Crooks-Jarzynski relation will help us gain more understanding of processes between non-equilibrium states will be a question we wish to address in the future.

## ACKNOWLEDGMENTS

I wish to thank Michele Bock for constant support and encouragement and Richard F. Martin, Jr. and other members of the physics department at Illinois State University for creating a supportive academic environment. I also wish to thank the three reviewers of this article for making very helpful and constructive comments and suggestions, by which the article has been greatly improved.

# REFERENCES

1. D. J. Evans, E. G. D. Cohen, and G. P. Morriss, *Probability of Second Law Violations in Shearing Steady States*, Phys. Rev. Lett. **71**, 2401 (1993).

2. D. J. Evans and D. J. Searles, *Equilibrium microstates which generate second law violating steady states*, Phys. Rev. E **50**, 1645 (1994).

3. G. Gallavotti and E. G. D. Cohen, *Dynamical Ensembles in Nonequilibrium Statistical Mechanics*, Phys. Rev. Lett. **74**, 2694 (1995).

4. G. Gallavotti, *Extension of Onsager's reciprocity to large fields and chaotic hypothesis*, Phys. Rev. Lett. **77**, 4334 (1996).

5. J. Kurchan, *Fluctuation theorem for stochastic dynamics*, J. Phys. A **31**, 3719 (1998).

6. G. E. Crooks, *Entropy production fluctuation theorem and the nonequilibrium work relation for free energy difference*, Phys. Rev. E **60**, 2721 (1999).

7. J. L. Lebowitz and H. Spohn, *A Gallavotti-Cohen-Type Symmetry in the Large Deviation Functional for Stochastic Dynamics*, J. Stat. Phys. **95**, 333 (1999).

8. C. Maes, *The Fluctuation Theorem as a Gibbs Property,* J. Stat. Phys. **95**, 367 (1999).

9. C. Jarzynski, *Hamiltonian derivation of a detailed fluctuation theorem*, J. Stat. Phys. **98**, 77 (2000).

10. B. Piechocinska, *Information erasure*, Phys. Rev. A **61**, 062314 (2000).

11. C. Maes, *On the Origin and the Use of Fluctuation relations for the Entropy*, Sém. Poincaré **2**, 29 (2003).

12. U. Seifert, *Entropy Production along a Stochastic Trajectory and an Integral Fluctuation Theorem*, Phys. Rev. Lett. **95**, 040602 (2005).

13. C. Jarzynski, *Nonequilibrium Equality for Free Energy Differences*, Phys. Rev. Lett. **78**, 2690 (1997).

14. M. Esposito, U. Harbola, and S. Mukamel, *Nonequilibrium fluctuations, fluctuation theorems, and counting statistics in quantum systems*, Rev. Mod. Phys. **81,** 1665 (2009).

15. M. Campisi, P. Hänggi, and P. Talkner, *Quantum Fluctuation Relations: Foundations and Applications*, Rev. Mod. Phys. **83**, 771 (2011).

16. G. E. Crooks, *Path-ensemble averages in systems driven far from equilibrium*, Phys. Rev. E **61**, 2361 (2000).

17. T. Yamada and K. Kawasaki, *Nonlinear Effects in the Shear Viscosity of Critical Mixtures*, Prog. Theor. Phys. **38**, 1031 (1967).

18. K. Kawasaki and J. D. Gunton, *Theory of Nonlinear Transport Processes: Nonlinear Shear Viscosity and Normal Stress Effects*, Phys. Rev. A **8**, 2048 (1973).

19. G. P. Morriss and D. J. Evans, *Isothermal Response Theory*, Mol. Phys. **54**, 629 (1985).

20. D. J. Evans and D. J. Searles, *Steady states, invariant measures, and response theory*, Phys. Rev. E **52**, 5839 (1995).

21. K. Hayashi and S. Sasa, *Linear response theory in stochastic many-body systems revisited*, Physica A **370**, 407 (2006).

22. R. Kubo, *Statistical-mechanical theory of irreversible processes. I*, J. Phys. Soc. Jpn. **12**, 570 (1957). L. Onsager, *Reciprocal relations in irreversible processes. I*, J. Phys. Rev. **37**, 405 (1931). R. Kubo, M. Toda, and N. Hashitsume, *Statistical Physics II: Nonequilibrium Statistical Mechanics* (Springer-Verlag, Berlin, 1991), Ch. 4.

23. G. N. Bochkov and Yu. E. Kuzovlev, *General theory of thermal fluctuations in nonlinear systems*, Sov. Phys. JETP **45**, 125 (1977).

24. D. Andrieux and P. Gaspard, *Quantum Work Relations and Response Theory*, Phys. Rev. Lett. **100**, 230404 (2008).

25. J. Kurchan, *A Quantum Fluctuation Theorem*, arXiv: cond-mat/0007360.

26. H. Tasaki, *Jarzynski relations for quantum systems and some applications*, unpublished note (2000), arXiv: cond-mat/0009244.

27. K. Saito and A. Dhar, *Fluctuation Theorem in Quantum Heat Conduction,* Phys. Rev. Lett. **99**, 180601(2007).

28. K. Saito and Y. Utsumi, *Symmetry in full counting statistics, fluctuation theorem, and relations among nonlinear transport coefficients in the presence of a magnetic field,* Phys. Rev. B **78**,115429(2008).

29. D. Andrieux, P. Gaspard, T. Monnai, and S. Tasaki, *Fluctuation theorem for currents in open quantum systems*, New J. Phys. **11**, 043014 (2009).

30. H. Matsuoka, *Green-Kubo formulas with symmetrized correlation functions for quantum systems in steady states: the shear viscosity of a fluid in a steady shear flow*, J. Stat. Phys. **148**, 933 (2012).

31. P. Talkner, P. Hänggi, and M. Morillo, *Microcanonical quantum fluctuation theorems*, Phys. Rev. E **77**, 051131 (2008).